\documentclass[manuscript]{aastex}

\shorttitle{Grain Alignment in Starless Cores}
\shortauthors{Jones et al.}

\begin{document}

\title{Grain Alignment in Starless Cores}

\author{T. J. Jones~\altaffilmark{1}, M. Bagley~\altaffilmark{1}}
\affil{Minnesota Institute for Astrophysics, University of Minnesota, Minneapolis, MN 55455}
\email{tjj@astro.umn.edu}

\author{M. Krejny~\altaffilmark{1}}
\affil{Cree Inc., 4600 Silicon Dr., Durham, NC}

\author{B-G Andersson}
\affil{SOFIA Science Center, USRA, Moffett Field, CA}

\author{P. Bastien}
\affil{Centre de recherche en astrophysique du Qu\'ebec and D\'epartment de Physique, Universit\'e de Montr\'eal, Montr\'eal, Canada}

\altaffiltext{1}{Visiting Astronomer at the Infrared Telescope Facility which is operated by the University of Hawaii under contract from the National Aeronautics and Space Administration.}

\begin{abstract}

We present near infrared polarimetry data of background stars shining through a selection of starless cores taken in the $K$ band, probing visual extinctions up to $A_{V} \sim 48$.  We find that $P_K/{\tau _K}$ continues to decline with increasing $A_{V}$ with a power law slope of roughly -0.5. Examination of published submillimeter (submm) polarimetry of starless cores suggests that by $A_{V} \gtrsim 20$ the slope for $P$ vs. $\tau$ becomes $\sim -1$, indicating no grain alignment at greater  optical depths. Combining these two data sets, we find good evidence that, in the absence of a central illuminating source, the dust grains in dense molecular cloud cores with no internal radiation source cease to become aligned with the local magnetic field at optical depths greater than $A_V \sim 20$. A simple model relating the alignment efficiency to the optical depth into the cloud reproduces the observations well.

\end{abstract}

\keywords{ISM: Magnetic Fields, ISM: Dust, ISM: Clouds}

\section{Introduction}

Dust grain alignment in protostellar environments is closely linked to both magnetic fields and the physical parameters of the dust grains in these regions. Since the discovery of polarized starlight by \citet{hiltner49} and \citet{hall49},  interstellar polarization has provided clues about the role that magnetic fields play in star formation \citep{crutcher04a}. The polarization caused by transmission through a medium with asymmetrical, aligned dust grains can be directly related to the orientation of the magnetic field in the plane of the sky.  Also, polarization (and extinction) measurements teach us about the size distribution, shape and structure of the dust grains themselves \citep{kim94a, kim94b}.

We know from observations that spinning, asymmetric dust grains align with their long axes perpendicular to the surrounding magnetic field \citep{jones96, lazarian03}. In fact, \citet{hiltner49} first suggested this connection. Depending on whether the polarization is in transmission or emission, the observed polarization is either parallel or perpendicular to the magnetic field, respectively.  However, despite years of research and theoretical work, we still do not have a comprehensive model of how the dust grains are actually aligned.   

Evidence is strong that in the very diffuse ISM grain alignment is nearly perfect. Computations using infinite cylinders by \cite{mathis86} reproduce the wavelength dependence of the polarization assuming perfect alignment of the large silicate grains. Using the observed trend of polarization with extinction in the optical and the wavelength dependence of the polarization, \cite{kim95} found that a high alignment fraction was necessary for the large silicate grains to produce the observed $P/\tau$, the polarization per unit optical depth. This was confirmed by \cite{draine09}.

There are three main grain alignment theories that have been developed to explain interstellar polarization. \cite{gold52} suggested that mechanical effects due to gas flow relative to the heavier grains could align grains, but this mechanism is not supported by the vast majority of observations \citep[e.g.,][]{hoang09}. \cite{davis51} developed a mechanism that relied on dissipation of rotational energy into heat in rotating grains through paramagnetic relaxation to align the grains with the magnetic field. Even with enhancements \citep[e.g.,][]{purcell79}, this theory is difficult to reconcile with observations \citep[e.g.,][]{roberge96, jones96}. Radiative torque alignment, originally proposed  by \cite{dolginov76}, has been explored more recently by various authors, including \cite{draine96, draine97}; \cite{lazarian07, lazarian11}; and \cite{hoang09, hoang08}. Recent work on observational tests for the radiative alignment theory can be found in \cite{whittet08}, \cite{andersson10, andersson11a}, and \cite{andersson11b}. Note that radiative alignment occurs with long grain axes perpendicular to B, i.e., in the expected orientation. This results not from paramagnetic relaxation but from averaging of radiative torques as the grain angular momentum precesses about B \citep{lazarian07}.

If grains were perfectly aligned everywhere, in all environments, then the grain alignment mechanism would be an interesting physics problem, but of only moderate importance to the study of the ISM using interstellar polarization. However, there is increasing evidence that grain alignment suffers dramatically reduced efficiencies in certain regions of the ISM, in particular dense molecular gas with no internal stars. Since these regions are sites of future star formation, and interstellar polarization provides an important tool for studying the magnetic field geometry in these regions, a realistic grain alignment model that accounts for these changes is a necessity.

Starless cores provide an important testbed for grain alignment theories, as optical depths can range beyond $A_{v} \sim 40$ \citep{sadavoy10}.  Unfortunately, because of these high optical depths, it is impossible to find stars bright enough to be seen at visual wavelengths behind these very dense  regions.  By obtaining polarization measurements at wavelengths from the Near-IR though the submm, we can work at these very high visual optical depths.  The fractional polarization as a function of extinction for various wavelengths can provide clues about grain alignment along these heavily extinguished lines of sight. 

The grain alignment efficiency has been measured primarily by studying the fractional polarization divided by optical depth ($P/\tau$) in both diffuse and dense regions. \citet{goodman92} observed that this ratio in dark clouds was significantly lower than in the diffuse ISM when viewing background field stars.  \citet{goodman95} and \citet{arce98} later suggested that beyond $A_{v} \sim$ a few, the grains in cold dark clouds could not polarize background field stars at all.  However, a different data set presented by \citet{gerakines95} showed that for background field stars, $P/\tau$ \textit{decreases} with extinction, but does \textit{not} go to zero.  In fact, an extensive analysis by \citet{whittet08} shows that $P/\tau$ decreases roughly as $\tau^{-0.5}$ in most dark clouds, but does not fall as $\tau^{-1}$, which would indicate the complete absence of grain alignment at the higher optical depths. 

\cite{clemens12} used $H$ band polarimetry results using Mimir \citep{clemens07} to study the magnetic field geometry and fractional polarization in a study of the L183 Starless Core. He finds that compared to the relatively unextincted surrounding stars, the reddened stars show no increase in polarization with extinction, suggesting that all of the polarization is induced in the outer layers of the cloud. These results probe extinctions up to $A_V \sim 14$ and do not penetrate the cloud core. \cite{clemens12} does, however, find clear evidence for an influence from the presence of the inner core of L183 on the magnetic field in the outer regions. \citet{alves14} studied the optical, Near-Infrared and submm polarimety of a starless core in the Pipe Nebula. They find that $P/\tau$ for the Near-Infrared polarimetry initially drops with a slope of $\tau^{-1}$, but beyond $A_V\sim10$ becomes more shallow with a slope of $\tau^{-.34}$. 

Polarization observations along lines of sight to embedded young stellar objects (YSOs), however, complicate the picture.  For these stars, the steep decline in $P/\tau$ with increasing $A_{V}$ is not seen \citep{jones89, whittet08}.  For embedded YSOs, the trend in $P/\tau$ can be explained by assuming complete grain alignment, but with the inclusion of turbulence in the magnetic field geometry \citep{jones92, jones89}. In other words, the grains along a line of sight to an embedded YSO appear to be well aligned, but not so along a line of sight to a background field star seen through molecular cloud material with no internal source of optical or Near-IR radiation. 

In this paper, we present $K$ band polarization and $J, H, K$ photometry of background stars extincted by starless cores.  Compared to previous measurements at somewhat lower $A_{V}$ \citep{whittet08, clemens12}, we see evidence for a further decrease in grain alignment for $A_{V} \ga 20$. Combining these results with an analysis of submm polarization data compiled from the SCUBA legacy project there is good evidence that grain alignment drops to zero at optical depths greater than $A_V \sim 20$.    

\section{Data}\label{sec-data}

\subsection{Source Selection for Observations}\label{ssec-omm}

Most of the molecular cores selected for NIR polarization observations were chosen for their high submm flux  \citep[L1544]{kirk05} or for their pre-existing submm polarization maps, such as CB3 \citep{ward-thompson09}, L43, L183, and L1544 \citep{ward-thompson00, crutcher04b}.  These submm observations allow us to make comparisons of the polarization at different wavelengths ($K$ band in extinction vs $850~\mu m$ in emission) and high values for $A_V$. A list of observing runs is given in Table 1.

\begin{deluxetable}{cccc}
\tablecaption{Summary of Observations}
\tablenum{1}
\label{table-obs}

\tablehead{\colhead{Cloud} & \colhead{Date Observed} & \colhead{Facility} & \colhead{Observation}}

\startdata
B133 & 23-25 Apr 2010 & OMM & JHK Phot\\
 & 21 May 2010 & OMM & JHK Phot\\
 & 3-5 Jun 2013 & IRTF & K Pol\\
B361 & 3-5 Jun 2013 & IRTF & K Pol\\
CB3 &10-13 Nov 2009 & OMM & JHK Phot\\
 & 10-12 Jan 2010 & IRTF &  K Pol\\
L1521 & 10-13 Nov 2009 & OMM & JHK Phot\\
 & 10-12 Jan 2010 & IRTF & K Pol\\
 & 3-5 Jun 2013 & IRTF & K Pol\\
L1544 & 10-13 Nov 2009 & OMM & JHK Phot\\
 & 10-12 Jan 2010 & IRTF & K pol\\
L183 & 11-16 Mar 2010 & OMM &  JHK Phot\\
 & 3-5 Jul 2010 & IRTF &  K Pol\\
 & 3-5 Jun 2013 & IRTF & K Pol\\
L43 & 3-5 Jul 2010 & IRTF & K Pol\\
\enddata

\end{deluxetable}

\subsection{Observations}

\subsubsection{NIR Photometry}

The imaging observations were obtained at the Mont-M\'egantic Observatory (OMM)
with the 1.6 m Ritchey-Chr\'etien telescope and the CPAPIR Near IR camera
\citep{artigau04}. The data were obtained in the $J$, $H$, and $K_{\rm
s}$ bands in queue observing mode (Artigau et al. 2010) during the periods
2009 November 10 - 13 (CB3, L1521, L1524, L1544, and L1498 [JH only]), and
2010 March 11 - 16 (L183),  April 23 - 25 and May 21 (B133). The field of
view is $\approx 30.3\arcmin$. The data reduction was performed with an in-house IDL program using the 2MASS PS Catalog for photometric and astrometric calibration. Table 2 lists the $JHK$ colors for (only) those sources with measurable polarization. 

Typical errors for the OMM photometry are $\pm 0.03$ mag. for $K<16$ and $H<16.5$. These photometric errors are much less than the errors in fractional polarization and conversion of $H-K$ to $A_V$, and are not given in Table 2. Note that the OMM field of view ($\sim 25\arcmin$) is far larger than the angular size of the target starless cores ($\sim 1\arcmin$), and we used the photometry only for sources that were very red and close to the center of the starless core (see end of \ref{pe}).

\subsubsection{IRTF Polarimetry}

Observations were taken at the Infrared Telescope Facility (IRTF) using NSFCAM2 \citep{shure94} in polarimetry mode \citep{jones97}.  Observations were taken during the nights of January 10-12 2010, July 3-5 2010, June 3-5 2013. Source selection was made based on a combination of red $H-K$ colors and $K$ magnitudes taken from the photometry described in the previous section.  All polarimetry observations were taken in the $K$ band.  For the polarimetry measurements, a half-wave plate was stepped to four different positions:  $0^{\circ}$ and $45^{\circ}$ corresponding to Stokes $Q$, and $22.5^{\circ}$ and $67.5^{\circ}$ corresponding to Stokes $U$.  The data analysis procedure is described in \citet{jones97}.  We had a seeing of $0.9^{\prime\prime}$, FWHM during the observations. All of the reduced polarimetry was corrected for polarization bias \citep{wardle74}.

Three types of calibrators were observed.  We observed BN and AFGL2591 to calibrate the polarization position angle.  NGC4147 and M75 were observed to calibrate the instrumental polarization ($P_{I}$) across the entire field of view.  The $P_{I}$ measurements were complemented by observing unpolarized standard stars HD90508, HD20194, and HD203856.  A list of observing runs is given in Table 1 and the results of our observations are summarized in Table 2 for those stars with measurable fractional polarization at $K$ ($P/\sigma_P > 3$). The column labeled $\epsilon P$ is the error in $P$ after correction for polarization bias. The conversion between $A_V$ and $\tau_K$ is given in Section \ref{results}.

Both stars in our $K$ band  observations in L183 are present in the $H$ band data set from \cite{clemens12}. The position angles agree nicely for both stars, but for star \# 21 in Clemens' list, our $H-K$ color is a bit redder and the fractional polarization at $K$ a bit higher than expected.

\subsection{Other Data}

We include data from three other sources in Table 2.  First, we include $K$ band polarization of the Bok Globules B118 and B361, taken from \citet{klebe90}. We have combined our new observations of B133 with results from \citet{klebe90}.  The one source in B335 was observed similarly to that of \citet{klebe90}, but was previously unpublished, and is presented here for the first time. For those sources not observed with OMM, colors were obtained from \citet{klebe90} or 2MASS \citep{skrutskie06}. Finally, for the single source in L1544, we obtained the $J$ magnitude from deep imaging using LUCI on the LBT \citep{seifert10}.

\section{Results}\label{results}

Figure 1 plots $J-H$ color versus $H-K$ color for the stars in our sample that have corresponding (bias corrected) polarization measurements with $P/\sigma_P > 3$. The two solid lines in the plot indicate the nominal reddening line for main sequence spectral types from B to M \citep{alves98}. We note that several of the sources with very red colors lie below the nominal reddening line. It could be these stars are embedded YSO's with thermal dust shells. This is unlikely, since these particular molecular cores are not associated with star formation and show no X-Ray sources indicative of T-Tauri chromospheres (although the very high extinction probably precludes X-Ray detections). They are likely dusty field giants in the background. Examination of WISE \citep{wright10} Band 1 and Band 2 colors confirms that these stars have a steep rise in flux beyond $3~\mu m$ and it is likely that the $H-K$ color is contaminated by thermal emission. For these three sources, we use the $J-H$ color to determine the extinction through the cloud (see Notes in Table 2). Note that the reddest source extends our data set to $A_V \sim 48$.

\begin{figure}\label{fig-colors}
\epsscale{1.0}
\plotone{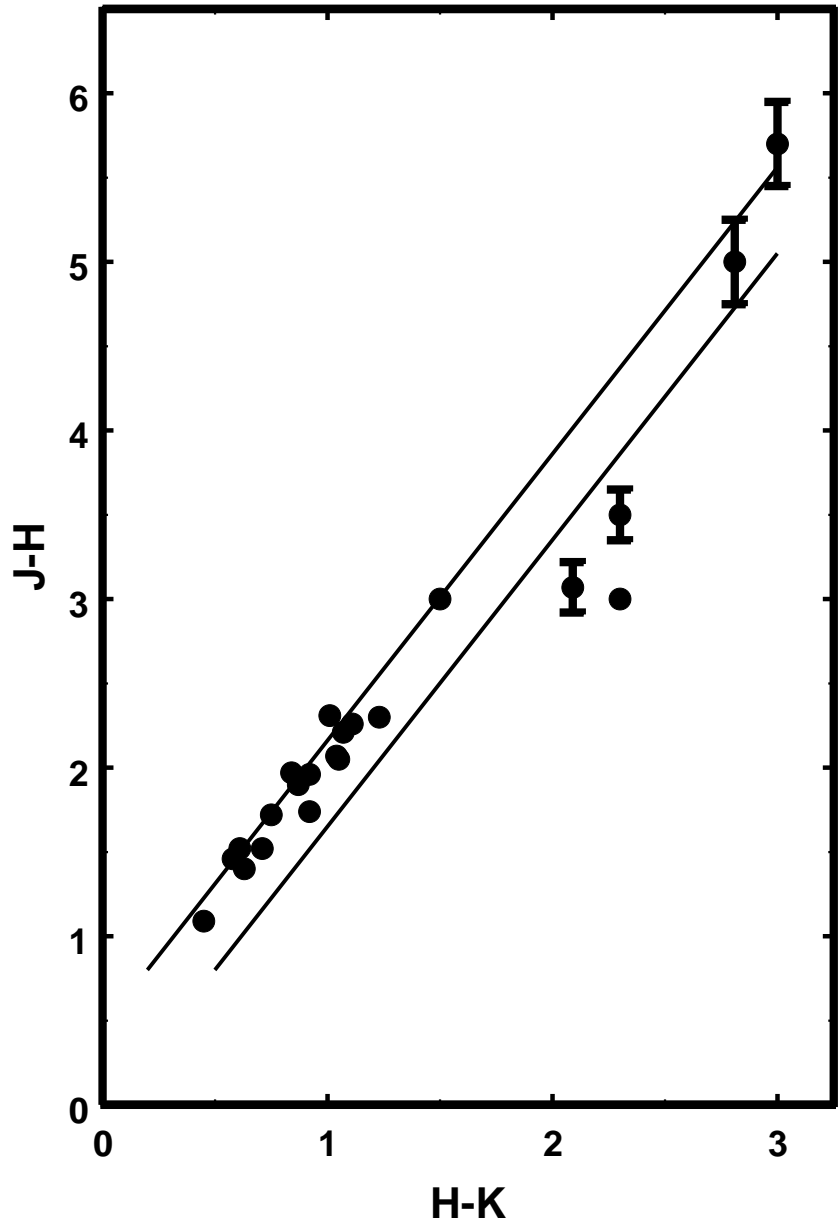}
\caption{Plot of of $J-H$ color versus $H-K$ color for the background stars for which we have obtained measurable $K$ band polarization. Error bars for $(J-H)$ are plotted only if they exceed $\pm0.1$ mag.}
\end{figure}

We compute the visual extinction $A_V$ from the $JHK$ colors using the formula in \cite{jones89, alves98}

\[\begin{array}{c}
{A_V} = 16.7E(H - K)\;{\rm{where}}\;E(H - K) = {(H - K)} - 0.1\\
{\rm{or}}\\
{A_V} = 10E(J - H)\;{\rm{where}}\;E(J - H) = {(J - H)} - 0.7\\
{\rm{and}}\\
{\tau_K} = 0.09A_V
\end{array}\]

Since we do not know the intrinsic colors of the stars, we have used typical values for $H-K$ colors. Since the intrinsic colors can range from $0<H-K<0.3$, an error of up to 0.2 magnitudes in $E(H-K)$ can be present, corresponding to an error of up to $A_V=3.3$ magnitudes. For the three stars where we were forced to use $E(J-H)$, the error in the color could be as much as 0.7 magnitudes, corresponding to an error of $A_V=7$ if the source is an early type star. Since early type stars are relatively rare in the field at Near-Infrared wavelengths \citep{jones81, garwood87} and less likely to have an infrared excess compared to dusty red giants, we have not tried to make any corrections to these sources. These three sources are plotted with a different symbol in Figures 2 and 5.

We will use the quantity $A_V$ to act as a proxy for $\tau_K$ since it is so commonly used as an extinction parameter in previous studies. We understand that there is evidence that the color excess $E(J - K)$ does not obey a simple linear correlation with the total dust column in lines of sight that intercept dense clouds \citep{whittet13}. However, we are ratioing the fractional polarization to the extinction at $K$, not the mass column density of dust along the line-of-sight. For the purpose of this analysis, we will assume the $E(H-K)$ color does linearly track the optical depth at $2.2~\mu m$ ($\tau_K$).

\subsection{Polarization Efficiency}\label{pe}

Figure 2 plots the fractional polarization per unit optical depth $(P/\tau)$ vs $A_{V}$ for all of the data sets described in Section \ref{sec-data} along with the observations from \citet{whittet08}.  These data are plotted with three computed trends. For polarization in transmission $P$ will increase with optical depth according to \citep{jones89}

\begin{equation}\label{Pmax}
P = \tanh ({\tau _P})
\end{equation}

\noindent where $\tau_P$ is proportional to the total optical depth, the proportionality depending on the grain alignment efficiency and the angle between the line-of-sight and the constant component of the magnetic field. The dot-dash line, labeled ${\rm{P}_{\rm{max}}}$ in Figure 2 corresponds to this equation for the case where the magnetic field lies in the plane of the sky (the optimum geometry) and grain alignment is maximum. The vertical location of ${\rm{P}_{\rm{max}}}$ is determined by the observed upper limit of interstellar polarization at $K$ \citep{jones89}, and is equivalent to the well know relation in the visible of ${\rm{P}}_{\rm{max}}=9E(B-V)$ from \cite{SMF}. The dip in ${\rm{P}_{\rm{max}}}$ at very high extinction is due to the fact that $P$ can not exceed 100\%, but the extinction can become arbitrarily large. The long-dash line is the JKD model for the trend of fractional polarization vs. optical depth at $K$ \citep{jones92}, which assumes perfect alignment but includes a turbulent component to the  magnetic field of equal strength to the constant component. This line well reproduces the mean trend found by \citet{whittet08} for embedded sources. 

The short-dash line corresponds to the fit given by \citet{whittet08} for field stars behind Taurus.  This fit of $P/\tau$ to extinction is given by a power law:  

\begin{equation}\label{pwrlaw}
\frac{{{P_K}}}{{{\tau _K}}} \propto {\left( {{A_V}} \right)^b}
\end{equation}

\noindent where Whittet finds  $b = -0.52 \pm 0.07$ for all of the field stars in his sample. We notice that, in this plot, our data at lower extinctions agree with the Whittet fit. At higher extinctions there is a suggestion of a further decrease in alignment efficiency at the highest optical depths.   

\begin{figure}\label{basic}
\epsscale{1.0}
\plotone{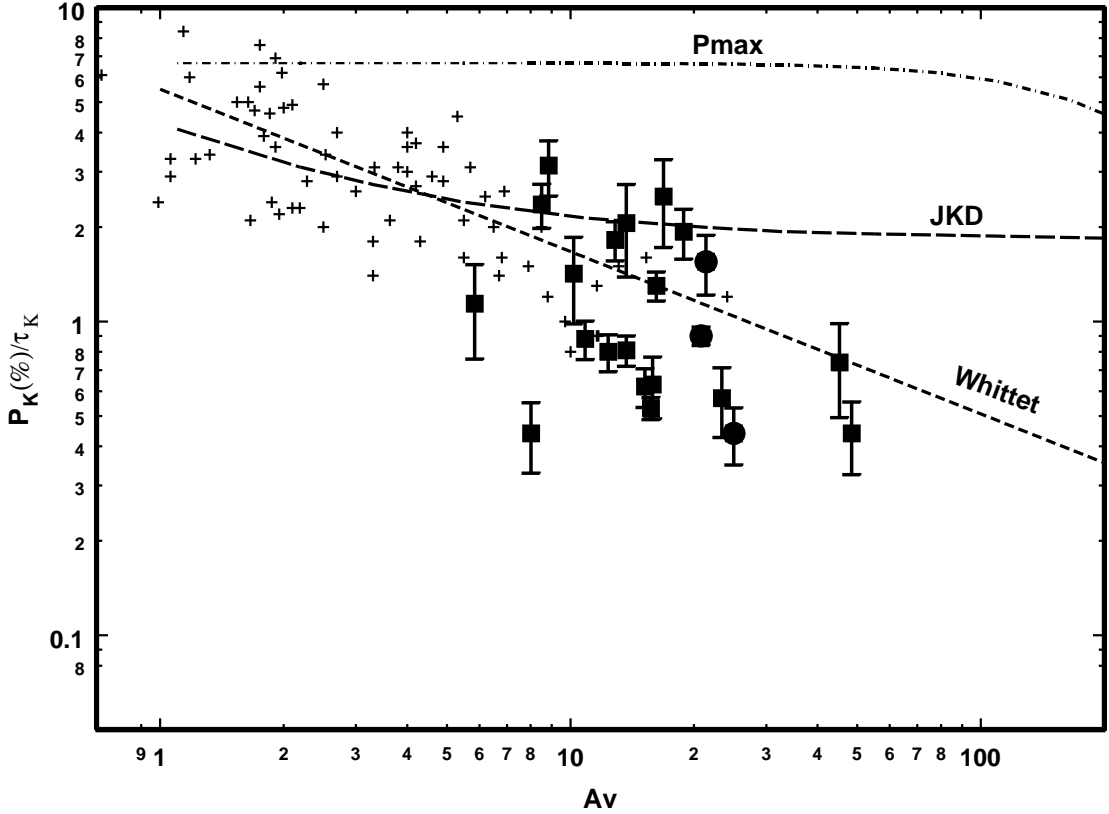}
\caption{Plot of of ratio $P_K/\tau_K$ versus extinction in $A_V$ for the background stars for which we have obtained measurable $K$ band polarization (solid squares and circles). The three sources for which $A_V$ was computed from $E(J-H)$ are plotted as solid circles. Also plotted are the background sources from \cite{whittet08} as plus signs. The dot-dash line labeled $\rm{P_{max}}$ is the case for perfect grain alignment with the magnetic field in the plane of the sky. The long-dash line labeled JKD includes turbulence \citep{jones89, jones92}. The short-dash line is from \cite{whittet08}.}
\end{figure}

We had hoped to examine the polarization of background stars at $K$ across the center of these starless cores, but selection effects have significantly reduced the available number of sources we could measure. Firstly, the extinction in the central core can easily exceed $A_V=50$, dimming the brightness of a star by 4.5 mag. or more at $K$. Secondly, the starless cores are typically $1 - 2\arcmin$ across, subtending a small solid angle on the sky, greatly reducing the odds of finding a bright field star shining through from behind. Since we have so few vectors, and none at the core center, we do not plot our position angles on a map of the sky. We do list the position angles for the sources in Table 2. On larger scales, \cite{klebe90}, \cite{clemens12}, \cite{jones84}, and \cite{ward-thompson09}, among others, present maps of the optical/Near-IR polarization vectors for the periphery of several starless cores.

\section{Comparison to Submm Data}

Considerable work has been done on modeling polarization in emission from molecular clouds. \cite{goncalves05} computed synthetic polarization maps for magnetized molecular clouds in order to determine the effects of magnetic field geometry on the fractional polarization that would be observed. They did not incorporate variations in grain alignment efficiency, but still found that a decrease in fractional polarization with column depth of gas could be explained by bending of magnetic field lines along the line of sight. \citet{padoan01} modeled polarized dust emission from starless cores that are assembled by supersonic turbulent flows. They incorporate the effects of grain alignment efficiency by turning off grain alignment beyond $A_V = 3$. \citet{pelkonen07, pelkonen09} studied the polarization of thermal dust emission to see if grain alignment by by radiative torques could explain the observed decline in the degree of polarization with mm and submm surface brightness. In their models the radiation necessary to align the grains is external to the molecular cloud, and they explore optical depths (column depth of the gas) up to about $A_V = 14$. 

\citet{bethell07} incorporate a radiative torque model for grain alignment into simulations of turbulent molecular clouds and compute synthetic polarization maps. In their model, the radiation necessary to align the grains also arises from outside the cloud. They find that the intensity and anisotropy of the intracloud radiation field show large variations throughout the models but are generally sufficient to drive widespread grain alignment at modest extinctions. The also find that the degree of polarization observed is extremely sensitive to the upper grain size cutoff, as expected for radiative torque alignment. Their largest column depths extend up to $A_V \sim 35$. \citet{alves14} find that the submm fractional polarization drops as $\tau^{-1}$ with flux ($A_V$) for the starless core Pipe-109 in the Pipe nebula. They model their results by assuming normal grain alignment up to a certain density, then having the grains become unaligned at higher density. In this way, the cloud is modeled as an outer shell with aligned grains and an inner core with unaligned grains. These studies generally find that grain alignment is minimal for $A_V > 10$ in the absence of a central source of radiation.

In this paper we seek to combine the observations of polarization in extinction with observations of polarized emission.  To facilitate this effort, we used the SCU-POL legacy data \citep{matthews09} to examine the trend of fractional polarization with submm surface brightness.  We selected only those points on the sky that had $P/\sigma_{P} > 3$.  These data were taken from the Vizier catalog.  However, the fluxes given in the table were given in volts and not calibrated.  We obtained calibrated flux measurements manually from the $850~\mu m$ polarization maps given in \citet{matthews09}.  We had the option of using the other references \citep{ward-thompson00, crutcher04b, kirk06,ward-thompson09}, but those only plotted the flux in terms of percent total flux instead of total flux.  We selected the two starless cores that had a sufficient number of $3\,\sigma$ polarization points to plot polarization versus extinction for the entire cloud.

We need to obtain $A_{V}$ estimates from the flux measurements given, and this calculation is presented in the Appendix I.  We note that this calculation is crude, and the final answer may be off by factors of 2 or more.  
Figures 3 and 4 plot the observed fractional polarization against Flux (or $A_V$) and the least squares power-law fit to the data. Fitting a power-law to $P$, similar to the power-law fit used for $P/\tau$ in extinction (eq. 2), we find slopes of $b = -1.19 \pm 0.24$ for L43 and $b = -1.18 \pm 0.15$ for L183. A slope of -1 indicates no grain alignment at these high optical depths. Although these data alone are not conclusive (and they represent only two cloud cores), the data do suggest that grain alignment turns off in these dense, starless cores at high $A_{V}$. \citet{poidevin10} present submm polarimetry of OMC-2 and OMC-3 where most of the cores are either Far Infrared sources or show evidence for an outflow, indicative of an embedded YSO. The region labeled 'South of FIR 6', however, appears to be quiescent and has 26 vectors. For this region we find a slope in $P$ vs. $\tau$ (Flux) of $-1.0 \pm 0.4$. Although not very well constrained, also consistent with an absence of grain alignment at large optical depths.

\begin{figure}\label{L183}
\epsscale{1.0}
\plotone{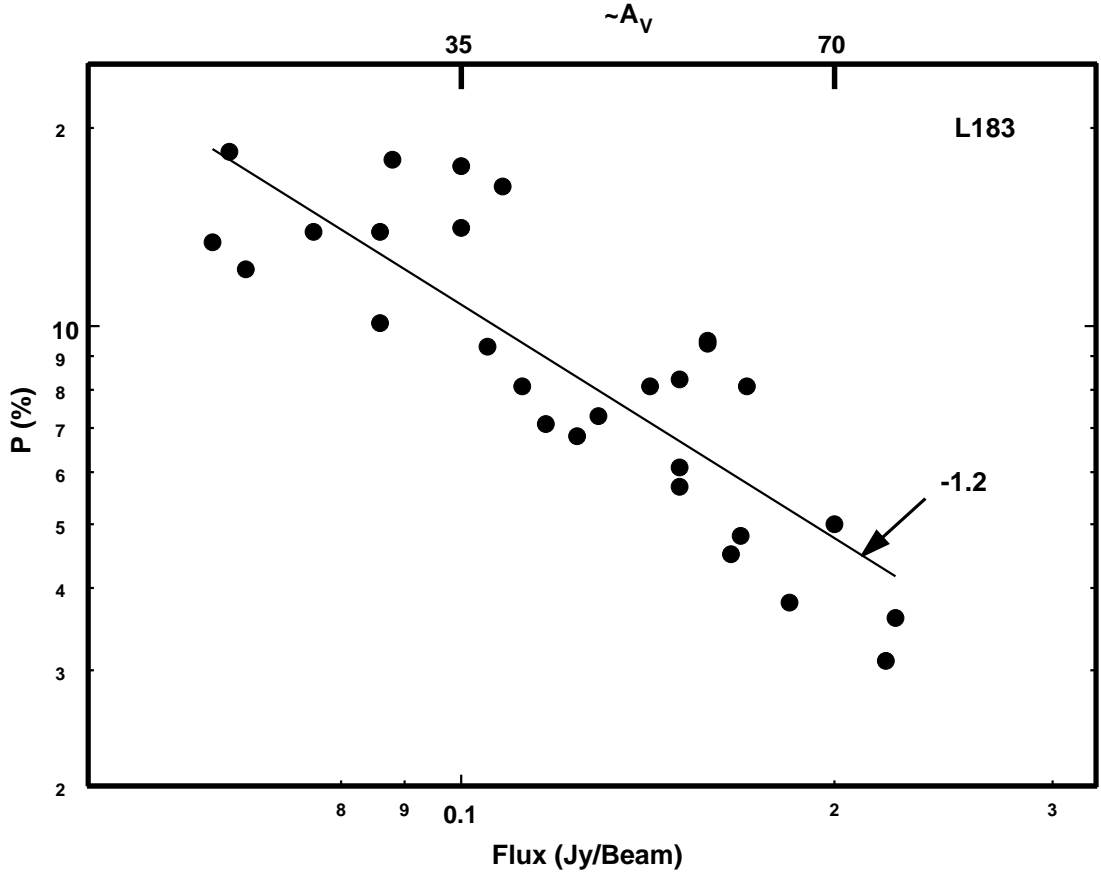}
\caption{Plot of fractional polarization at submm wavelengths of locations in the starless core L183 vs. surface brightness. An approximate equivalent $A_V$ is given along the top abscissa (see Appendix I).}
\end{figure}

\begin{figure}\label{L43}
\epsscale{1.0}
\plotone{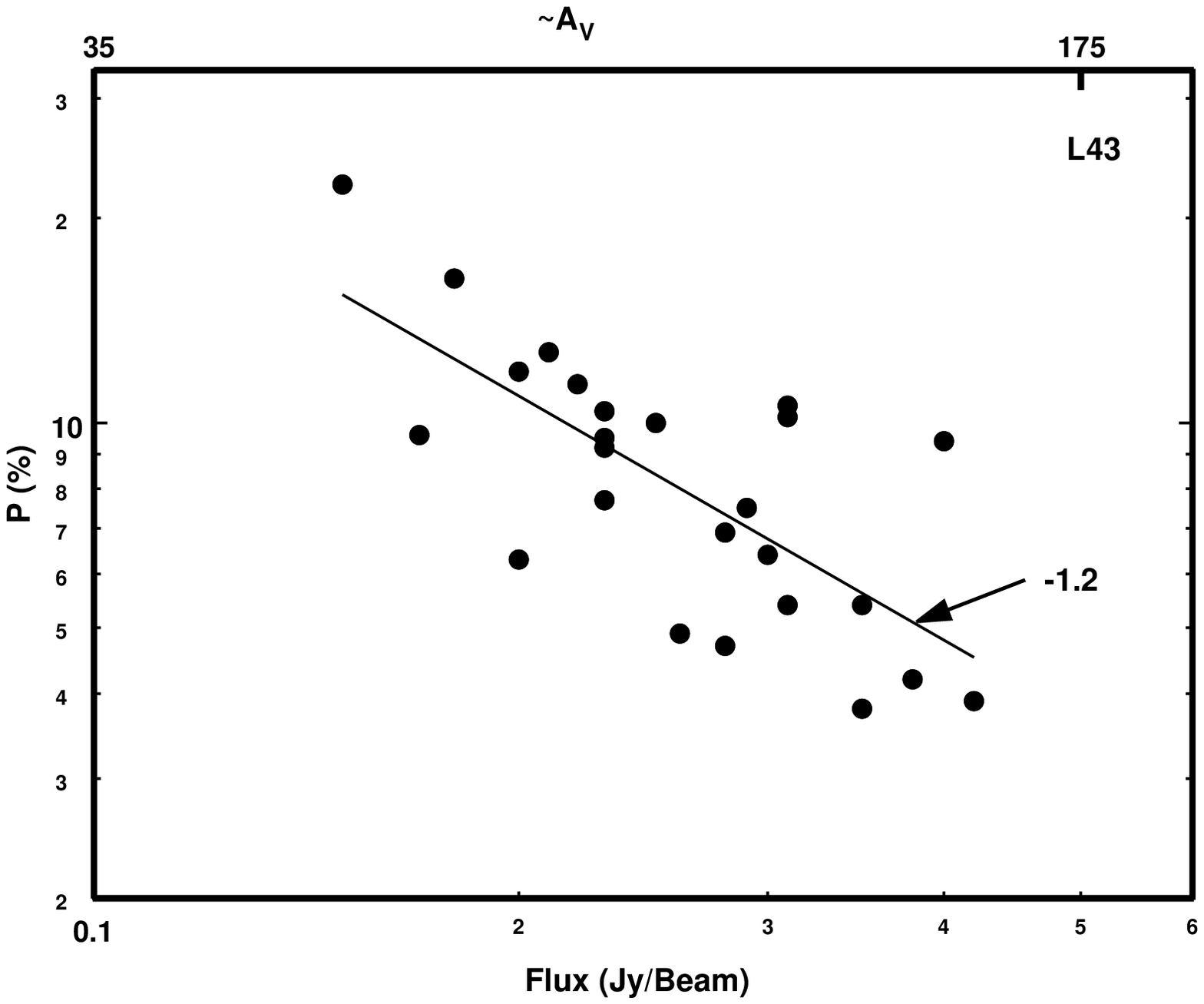}
\caption{Same as Figure 3, but for the L43 starless core.}
\end{figure}

We can plot the submm results on the same figure as the $K$ band polarimetry by using the conversion of submm flux to $A_V$ outlined in Appendix I and arbitrarily sliding the data points plotted in Figures 3 and 4 vertically to line up with the $K$ band extinction data in Figure 2. In Appendix II we provide a simple argument why this is a valid procedure. Although it may seem improper to directly compare polarization in transmission with polarization in emission, the trend with optical depth should be the same. If alignment is perfect with no turbulence, then the fractional polarization in emission will be constant with surface brightness (optical depth) provided the emission is optically thin, which is the case at submm wavelengths for all but the exact center for these starless cores. Thus, the submm data will plot (after vertical alignment) similar to ${\rm{P}_{\rm{max}}}$ in Figure 2. If there is a 50/50 mix of constant and random components to the magnetic field, then the polarization in emission should drop with optical depth through the first few decorrelation cells as $P \propto \tau ^{ - 0.5} $, then level out at a value determined by the constant component \citep[see Fig. 2 in][]{jones03}. This trend will closely follow the JKD model shown in Figure 2. Finally, if, only the outer skin of these molecular cloud cores has aligned grains, then the fractional polarization in emission will simply be diluted by added unpolarized flux (optical depth) as the line of sight penetrates the core, and we have $P \propto \tau^{-1}$, as seen in Figures 3 and 4. This is also case for $P/\tau$ in extinction, when the grains are no longer aligned, and increasing optical depth is not accompanied by increasing polarization.  Figure 5 shows the results of this comparison.

\begin{figure}\label{model}
\epsscale{1.0}
\plotone{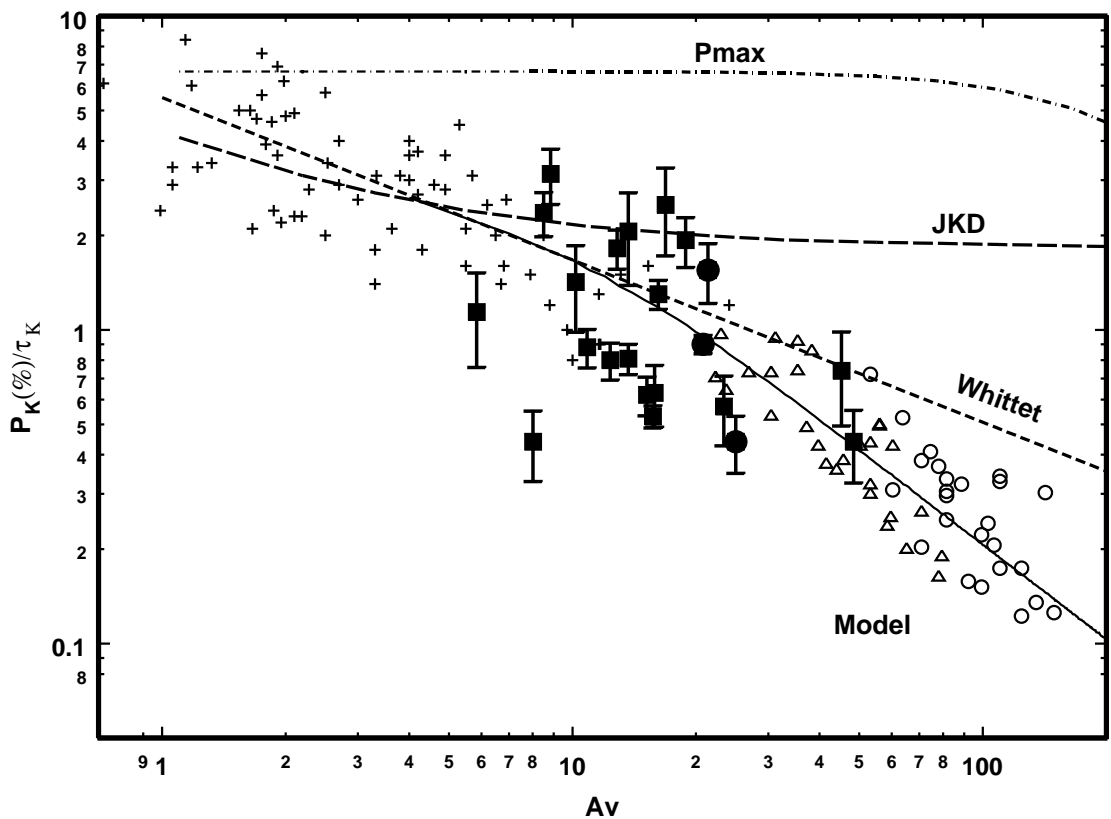}
\caption{Same as Figure 2, with the addition of submm observations and a simple model fit to the trend at high extinctions. The L183 data is plotted as open triangles and the L43 data as open circles. Each submm data set was shifted vertically by an arbitrary amount to line up with the $K$ band data (see Appendix II). The solid line is the result of a simple model where the alignment of the grains decreases as a function of optical depth into the cloud.}
\end{figure}

\section{A Simple Model}

\cite{whittet08} suggest that grain alignment by radiative torques appears to best explain the data. In radiative torque theory the alignment is sensitive to the ratio of wavelength to grain size ($\lambda / a$). Reddening of the external radiation field due to dust extinction progressively removes short wavelength photons which are necessary to align the smaller grains in the grain size distribution. With progressively fewer aligned grains at greater extinction, the value for $P/\tau$ drops. Using radiative torque alignment theory, Whittet et al. find this attenuation of the external radiation field appears adequate to account for the observed polarization in starless regions for extinctions up to $A_V \sim 10$ mag., the limit of their calculations. This study did not include the effects of turbulence.

Combining these results and the trend seen in Figure 5, there is now a strong indication that grains in at least some starless cores can become completely unaligned at very high extinctions ($A_V>20$), corresponding to the region deep in their interiors. The solid line in Figure 5 is a simple model that modifies the JKD model to vary the alignment efficiency (fraction of grains aligned) with optical depth. Using the notation in Jones (1989), the alignment efficiency is parametrized by $\eta$, where

\begin{equation}\label{eta}
\eta  = \frac{{{\kappa _ \bot }}}{{{\kappa _\parallel }}}\quad \rm{so}\quad {\tau _P} = \frac{{1 - \eta }}{{1 + \eta }}\tau 
\end{equation}

\noindent In this scheme $\kappa_\parallel$ and $\kappa_\bot$ are understood to be the total extinction coefficients with contribution from all sizes of dust grains. If no grains are aligned, then $\kappa_\parallel = \kappa_\bot$, $\eta = 1$ and the dust can not polarize starlight.

The nominal value of $\eta$ is 0.875, set by the maximum observed trend of $P_K$ with $\tau_K$ (Jones 1989, 1992) and is equivalent to the well known relation in the visible of ${\rm{P}}_{\rm{max}}=9E(B-V)$ from \cite{SMF}. A smaller $\eta$ would correspond to a higher polarizing power (not observed) and an $\eta$ closer to 1 corresponds to weaker polarizing power, or equivalently, fewer aligned grains.  In our modification of the JKD model, at each step in $\Delta\tau$ through the cloud the value of $\eta$ is no longer a constant, but is parametrized by

\[\begin{array}{l}\label{etatau}
{\tau _K} < 0.3\quad \eta ({\tau _K}) = {\eta _0}\\
{\tau _K} \ge 0.3\quad \eta ({\tau _K}) = 1 - {e^{ - 1.5(\tau _K-0.3)}}\left( {1 - {\eta _0}} \right)
\end{array}\]

\noindent where $\eta_0 = 0.875$. In other words, $\eta$ at low optical depths is 0.875 (maximum alignment), but increases to 1.0 at increasing optical depth (no alignment). The term controlling the decrease in efficiency is the extinction of light from the external radiation field (the $e^{-\tau}$ factor). The free parameters are the optical depth at which $\eta$ begins to decrease ($\tau_K=0.3$) and the factor of 1.5 in the exponent. The results of these model calculations produce the solid line in Figure 5. Note that this model includes the effects of turbulence, which accounts for the initial drop in $P/\tau$ for $A_V \lesssim 4$. In this case we are using the same parameters found in JKD, a 50/50 mix of the random and constant components with a decorrelation length of $\Delta\tau_K = 0.1$ for the random component. The only modification is the variation of $\eta$ with optical depth. 

The JKD model is a very simplistic parametrization of the physical conditions in the ISM in comparison to the models for submm polarized emission discussed at the beginning of this section,  but reproduces the main characteristics of the observations well. A simple decrease in alignment efficiency with optical depth can easily reproduce the observations both in extinction and in emission. This is a strong indication that reddening of the radiation field is very likely responsible for the decrease in $P/\tau$ with optical depth in quiescent molecular clouds.

The break point in the trend of $P/\tau$ from a slope of $\tau^{-0.5}$ to $\tau^{-1.0}$ is around $A_V\sim 20$, and this makes sense when considering radiative torque theory. We can ask, at what $A_V$ does the radiation that can couple to the largest grains disappear due to reddening?  To find this wavelength, roughly set $\tau_\lambda=1$ and use the radiative torque condition that $\lambda<2a$ (where $a$ is the grain radius).  Based on \citet{kim95}, the largest grains are $a\sim 2~\mu m$ and this implies $\lambda_{RT} \sim 4 \mu m$. If $A_V=20$, using the interstellar extinction curve given in \citet{rieke85} we find $A_V/A_\lambda=20$ at $\lambda = 3.7 \mu m$, or $\tau_\lambda=1$ at $\lambda \sim 3.6 \mu m$.

We note that the all sky Planck polarization data \citep{ade14} also shows a steep decline in \textit{mean} fractional polarization with 353 GHz surface brightness (translated to $n_H$) for neutral hydrogen column depths $2\times10^{22}~\rm{cm}^{-2}<N_H<4\times10^{22}~\rm{cm}^{-2}$. This translates to $15<A_V<30$ (using $N_H/A_V=1.25\times 10^{21}~\rm{cm}^{-2}$), which spans the range in $A_V$ for which we see the break in the slope in Figure 5. It is difficult to directly compare the Planck results to ours, given that the Planck beam of $5\arcmin$ does not resolve starless cores, but the slope of the \textit{mean} polarization with optical depth ($P/\tau$) is $\sim -1$ over this range in $A_V$, consistent with the absence of grain alignment for $A_V>20$.

\section{Conclusions}

In this paper, we presented new $K$ band polarimetry of background field stars extincted by dense starless cores.  Combining our new data with that previously obtained, we found basic agreement with the power law used by \citet{whittet08} to describe the trend of $P/\tau$ with extinction $A_V$.  Submm plots of the percent polarization versus flux/extinction for individual clouds show slopes near $\tau^{-1}$, indicative of no grain alignment. When the Near IR observations are combined with the Submm observations, there is a strong indication that grain alignment decreases to zero deep inside molecular cloud cores with no internal source of radiation (i.e. an embedded YSO). A simple model invoking a decrease in alignment efficiency with optical depth can easily reproduce the observations. There is a change in slope from $\tau^{-0.5}$ to $\tau^{-1}$ at $A_V\sim20$, as predicted by the Radiative Torque model for grain alignment.

\section {Acknowledgements}

We would like thank the IRTF staff and Kathleen DeWahl for assistance with the observations. We also  thank the technicians of the Observatoire du Mont-M\'egantic (OMM) and the students who took the images in queue mode. This Observatory was supported in part by a Natural Sciences and Engineering Council (Canada) grant. This research has also been supported in part by National Science Foundation grant AST-1109167.

\begin{center}
      {\bf APPENDIX I}
\end{center}

To calculate $A_{V}$ in terms of the $850\,\mu$m flux $F$, we first determine the submm optical depth $\tau(\nu)$ from the observed flux $F$ by assuming a dust temperature $T=10~K$ and using the blackbody equation:

\begin{equation}\label{eq-temp}
F(\nu,T) = B(\nu,T)\tau(\nu)\Omega
\end{equation}

\noindent where $B(\nu,T)$ is the blackbody function and $\Omega$ is the solid angle of the beam (for the published SCU-POL observations the beam is $\sim 20^{\prime\prime}$). 

We can estimate the hydrogen column density in terms of the optical depth $\tau(\nu)$ as given by \citet{hildebrand83}:  

\begin{equation}
N(H + H_2)/\tau = 1.2 \times 10^{25} \times (\lambda/400)^{2}\,atom\,cm^{-2}
\end{equation}\label{eq-hildebrand}

\noindent where $\lambda$ is the wavelength in $\mu$m (here, 850).  Note that a different equation is used for wavelengths below $400\,\mu$m and that we consider $N(H_{2}) = 0.5N(H)$.

Equation 7-18 of \citet{spitzer78}  relates the hydrogen column density ($N_{H}$) to the color excess:

\begin{equation}
N_{H} = 5.9 \times 10^{21} E(B-V) mag^{-1} cm^{-2}
\end{equation}\label{eq-spitzer718}

\noindent and the extinction in terms of color excess is given by \citep[Eq. 7-20]{spitzer78}

\begin{equation}\label{eq-sp}
A_{V} = R_{V} E(B-V)
\end{equation}

\noindent where $R_{V}$ is the ratio of general to selective extinction.  We use the value for the diffuse ISM ($R_{V} \sim 3$). 

Combining all this and solving for $A_{V}$ gives:

\begin{equation}
{A_V} = 2{R_V}\frac{{1.2 \times {{10}^{25}}}}{{5.9 \times {{10}^{21}}}}{\left( {\frac{\lambda }{{400}}} \right)^2}\tau 
\end{equation}

\begin{center}
      {\bf APPENDIX II}
\end{center}

Using the notation of \cite{jones89} and ignoring turbulence, we can write the net fractional polarization due to extinction as

\begin{equation}
P_{ex} = \tanh ({\tau _P})
\end{equation}

\noindent where $\tau_P$ is defined in (\ref{eta}) and in Jones (1989). If $\eta$ varies along the line-of-sight, we can write

\begin{equation}
{\tau _P} = \int_0^\tau  {\frac{{1 - \eta (\tau )}}{{1 + \eta (\tau )}}} d\tau
\end{equation}

If $\tau_P$ is small, as is the case for the few percent polarization we measure for background field stars, then $P \sim \tau_P$ and the 'efficiency' in the $K$ band is

\begin{equation}
\frac{{{P_{ex}}(K)}}{{\tau (K)}} \approx \frac{{{\tau _P}(K)}}{{\tau (K)}}
\end{equation}

\noindent Note that $\tau$ at $K$ can be very large, even if $\tau_P$ is not.

In emission we have, again, ignoring turbulence \citep{dennison77}

\begin{equation}
{I_\parallel } = \frac{B}{2}\left( {1 - {e^{ - {\tau _\parallel }}}} \right)\quad {\rm{and}}\quad {I_ \bot } = \frac{B}{2}\left( {1 - {e^{ - {\tau _ \bot }}}} \right)
\end{equation}

\noindent where $B$ is the blackbody function. Noting that $\tau_\parallel = \tau + \tau_P$ and $\tau_\bot = \tau - \tau_P$ we can express the polarization in emission by

\begin{equation}
{P_{em}} = \frac{{{I_\parallel } - {I_ \bot }}}{{{I_\parallel } + {I_ \bot }}} = \frac{{\sinh ({\tau _P}){e^{ - \tau }}}}{{1 - \cosh ({\tau _P}){e^{ - \tau }}}}
\end{equation}

\noindent If the emission is optically thin at Submm wavelengths (both $\tau$ and $\tau_P$ are small), we have

\begin{equation}
{P_{em}}({\rm{Submm}}) \sim \frac{{{\tau _P}(1 - \tau )}}{{1 - (1 - \tau )}} \sim \frac{{{\tau _P}({\rm{Submm}})}}{{\tau ({\rm{Submm}})}} \propto \frac{{{\tau _P}(K)}}{{\tau (K)}}
\end{equation}

\noindent The proportionality on the right is not an equality because the extinction efficiencies are wavelength dependent. Also, the proportionality on the right assumes that the grain population causing the polarization in extinction is the same is the grain population causing the polarization in emission. We recognize that this has not been established by observation, and is possibly incorrect along some lines of sight \citep{lazarian97}. If this assumption is valid for L183 and L43,  we can write

\begin{equation}
{P_{Submm}} \propto \frac{{{P_K}}}{{{\tau _K}}}
\end{equation}

Since we have already established the relationship between optical depths at Near-IR and Submm wavelengths in Appendix I, we use this proportionality to justify the vertical scaling of the Submm polarization in Figure 5. For L183, the scaling factor is $P_K/\tau_K=0.054P_{Submm}$ and for L43 the scaling factor is $P_K/\tau_K=0.032P_{Submm}$. These scaling factors are not equal due to differences in the angle the constant component makes to the line-of-sight, differences in the dust temperature used in (\ref{eq-temp}), or differences in other assumptions we have made.

\begin{deluxetable}{cccccccccccc}
\tablecaption{Data}
\tablenum{2}
\label{table-photometry}

\tablehead{\colhead{Cloud} & \colhead{RA} & \colhead{Dec} & \colhead{$K$} & \colhead{$H-K$} & \colhead{$J-H$} & \colhead{$P_{K}(\%)$} & \colhead{$\epsilon P$} & \colhead{$\theta$} & \colhead{$A_V$} & \colhead{$P/\tau_K$} & \colhead{Notes}}

\startdata
B118 & 18:53:58.0 & -7:26:04 & 10.9 & 0.71 & 1.52 & 1.3 & 0.4 & 22 & 10.2 & 1.4 & (2) \\
B118 & 18:53:57.4 & -7:26:03 & 11.2 & 0.63 & 2.08 & 2.5 & 0.5 & 18 & 8.9 & 3.1 & (2) \\
B118 & 18:53:54.9 & -7:25:55 & 11.7 & 1.05 & 2.4 & 0.9 & 0.2 & 130 & 15.9 & 0.63 & (2) \\
B133 & 19:06:08.6 & -6:53:15 & 12.9 & 2.8 & 5.0 & 3.0 & 1.0 & 16 & 45.3 & 0.74 &  \\
B133 & 19:06:16.1 & -6:53:31 & 10.8 & 2.3 & 3.5 & 1.1 & 0.25 & 176 & 25.0 & 0.44 & (1) \\
B133 & 19:06:10.2 & -6:54:04 & 10.2 & 1.5 & 3.0 & 1.2 & 0.3 & 35 & 23.4 & 0.57 & \\
B133 & 19:06:26.8 & -6:55:08 & 10.0 & 2.30 & 2.86 & 1.7 & 0.2 & 91 & 20.6 & 0.91 & (1) \\
B133 & 19:06:08.6 & -6:54:33 & 9.0 & 0.75 & 1.72 & 0.86 & 0.12 & 44 & 10.9 & 0.88 & \\
B133 & 19:06:17.1 & -6:52:23 & 6.1 & 0.92 & 1.74 & 1.0 & 0.11 & 171 & 13.7 & 0.81 & \\
B133 & 19:06:14.8 & -6:51:53 & 10.8 & 0.61 & 1.52 & 1.81 & 0.29 & 11 & 8.5 & 2.4 & \\
B133 & 19:06:08.6 & -6:53:56 & 12.6 & 1.11 & 2.26 & 3.79 & 1.18 & 102 & 16.9 & 2.5 & \\
B335 & 19:36:58.7 &  7:34:00 & 7.6 & 1.07 & 2.21 & 1.9 & 0.2 & 147 & 16.2 & 1.3 & (4) \\
B361 & 21:12:22.5 & 47:23:15 & 10.0 & 0.84 & 1.97 & 0.89 & 0.12 & 20 & 12.4 & 0.80 & (2) \\
B361 & 21:12:04.4 & 47:20:19 & 8.4 & 0.58 & 1.46 & 0.32 & 0.08 & 127 & 8.0 & 0.44 & (2) \\
B361 & 21:12:19.2 & 47:20:24 & 10.9 & 0.87 & 1.9 & 2.1 & 0.3 & 57 & 12.9 & 1.8 & (2) \\
B361 & 21:12:16.4 & 47:20:15 & 9.5 & 1.01 & 2.31 & 0.85 & 0.12 & 70 & 15.2 & 0.62 & (2) \\
B361 & 21:12:15.1 & 47:24:21 & 11.5 & 1.23 & 2.30 & 3.28 & 0.59 & 63 & 18.9 & 1.9 & (2) \\
B361 & 21:12:22.0 & 47:24:45 & 10.9 & 0.92 & 1.96 & 2.54 & 0.83 & 48.3 & 13.7 & 2.1 & (2) \\
CB3  & 00:28:45.2 & 56:43:07 & 12.9 & 2.09 & 3.07 & 3 & 1 & 44 & 21.4 & 1.6 & (1) \\
L1544 & 05:04:15.9 & 25:11:57 & 13.0 & 3.0 & 5.7 & 1.9 & 0.5 & 176 & 48.4 & 0.44 & (3) \\
L183 & 15:54:29.3 & -2:48:23 & 9.7 & 0.59 & 0.69 & 0.6 & 0.2 & 96 & 5.8 & 1.1 & \\
L183 & 15:54:20.4 & -2:54:07 & 7.9 & 1.04 & 2.07 & 0.75 & 0.06 & 75 & 15.7 & 0.53 & \\
\enddata

Notes: (1) $A_V$ from $E(J-H)$ (2) Klebe \& Jones (1990); (3) J from LBT; (4) unpub. \\

\end{deluxetable}

\end{document}